\documentclass[aps,pre,amsmath,amssymb,superscriptaddress,floatfix,showpacs,preprint]{revtex4}

\usepackage{graphicx}% Include figure files
\usepackage{bm}% bold math
\input epsf.sty

\begin{document}

\title{Entropy production in the non-equilibrium steady states of interacting many-body systems}

\author{Sven Dorosz}
\affiliation{Theory of Soft Condensed Matter, Universit\'{e} du Luxembourg, Luxembourg, L-1511  Luxembourg}
\affiliation{Department of Physics, Virginia Polytechnic Institute and State University, Blacksburg, Virginia 24061-0435, USA}

\author{Michel Pleimling}
\affiliation{Department of Physics, Virginia Polytechnic Institute and State University, Blacksburg, Virginia 24061-0435, USA}

\date{\today}

\begin{abstract}
Entropy production is one of the most important characteristics of non-equilibrium steady
states. We study here the steady-state entropy production, both at short times as well as in the long-time
limit, of two important classes of non-equilibrium systems: transport systems and reaction-diffusion systems.
The usefulness of the mean entropy production rate and of the
large deviation function of the entropy production for characterizing non-equilibrium steady states
of interacting many-body systems is discussed.
We show that the large deviation function 
displays a kink-like feature at zero entropy production
that is similar to that observed for a single particle driven along a periodic potential.
This kink is a direct consequence of the detailed fluctuation theorem fulfilled by the probability
distribution of the entropy production and is therefore a generic feature of the corresponding
large deviation function.
\end{abstract}

%\pacs{05.40.-a,05.70.Ln,05.20.-y}% PACS, the Physics and Astronomy

\maketitle
\pagestyle{plain}
\section{Introduction}
The study of large deviation functions has been of increasing importance for the understanding of many-body systems.
On the one hand large deviation functions form the basis of a modern approach to equilibrium statistical mechanics \cite{Ell85,Dem98,Tou09},
on the other hand they are increasingly recognized of being of fundamental interest for the characterization of non-equilibrium systems where they
are sometimes considered to play a role similar to that played by the free energy at equilibrium. Important examples are given
by the large deviations of the steady-state currents \cite{Ber07,Der07} in
systems that are far from equilibrium.

Recently, Mehl {\it et al.} \cite{Meh08} extended the study of large deviation functions far from equilibrium to
the steady-state entropy production. Studying a
one-dimensional system composed of a single particle driven along a periodic potential, they reformulated the problem
as a time-independent eigenvalue problem \cite{Leb99} and showed that for this system
the large deviation function of the entropy production exhibits a kink at zero entropy production. 
Similar kinks in the large deviation function of the entropy production or in related large deviation
functions of other quantities can also be found in a range of other systems \cite{Vis06,Tur07,Cle07,Lac08,Kum10,footnote1}.
These are intriguing results
that raise the important question whether the presence of this kink is a universal feature of systems with non-equilibrium
steady states.

In this paper we study the entropy production in the steady states of different non-equilibrium
interacting many-body systems. On the one hand we study the Partially Asymmetric Simple Exclusion Process (PASEP) \cite{Bly07},
a transport process in an open system where particles can enter or leave only at the system boundaries.
On the other hand we investigate reaction-diffusion systems on a ring where the number of particles can change everywhere in the system
due to reactions between the particles. 
Our main emphasis thereby is on the entropy production in the long-time limit where we study the large deviation function in the
same way as done by Mehl {\it et al.} in their study of the single particle system.
For all studied many-body systems we find a kink-like feature at zero entropy production, similar to what has been observed for the single
particle system.
We show that this kink is a generic feature in non-equilibrium systems obeying a detailed fluctuation theorem for the
entropy production.

It is worth noting that large deviation functions of the {\it current} have been studied previously in some related systems.
Studies of the Totally Asymmetric Simple Exclusion Process (TASEP) \cite{Der98,Der99} 
and of the Symmetric Simple Exclusion Process (SSEP) \cite{Kim95,Leb99} revealed a non-analytical behavior
of the current large deviation function at the value 0 of the parameter that is canonically conjugated
to the particle current. Using dynamical renormalization group techniques, it was later shown that the value of the corresponding
exponent follows from the noise renormalization \cite{Lec07}. Very recently, Bodineau and Lagouge \cite{Bod10} studied
the current large deviations in a driven dissipative lattice gas model with creation and annihilation processes, whereas
Simon \cite{Sim10} investigated the large deviation function of the current
for the Weakly Asymmetric Simple Exclusion Process on a ring.

Our paper is organized in the following way. In the next Section we introduce our models, before discussing in Section III
our approaches for computing the entropy production at short times as well as the mean entropy production
rate and the large deviation function for the entropy production
in the long-time limit. Our results are presented and discussed in Section IV. Finally, Section V gives
our conclusions.

\section{Models}
In order to elucidate the entropy production in interacting many-body systems, we discuss in the following reaction-diffusion
systems on a ring as well as open transport systems where particles move through a system that they can enter or leave 
only at its boundaries.
In the past, due to the combination of their conceptual simplicity and highly non-trivial results,
diffusion-limited reaction systems \cite{Hen08,Odo08} and simple exclusion processes \cite{Der98a,Sch01,Gol06}
have greatly contributed to our understanding
of processes far from equilibrium. For the same reasons, these models are also the natural choices for 
our study of the entropy production in many-body systems.

As reaction-diffusion systems we consider simple cases where particles $A$ diffuse on a one-dimensional ring, under
the condition that every lattice site can only be simply occupied. This diffusion process is mimicked by the hopping
of a particle to an empty nearest neighbor site with rate $D$. In addition, we also allow for particle creation and annihilation.
In the creation process a new particle can be created at an empty site with rate $h$, whereas in the annihilation 
process $n$ particles on $n$ connected sites are destroyed with rate $\lambda$, 
yielding $n$ empty sites. In the following we characterize
our models by the number of particles involved in the annihilation process and call $M_n$ (with $n = 2 ,3$) the model 
where $n$ particles are destroyed. As a variant, we also study the situation (we call the resulting model $M_{2'}$) where 
in a two particles reaction only one particle is destroyed.

These models are the same as those studied in \cite{Dor09,Dor09b,Dor10} in order
to better understand steady-state and transient properties of reaction-diffusion systems. 
All these systems have non-equilibrium steady states. In addition, some or all reactions do not allow
for a direct back-reaction, and microscopic reversibility is broken. However, as discussed in the next Section,
the computation of the entropy production requires microscopic
reversibility, i.e. for every reaction the direct back-reaction must be possible.
For that reason we are adding to the reaction schemes the direct back-reactions of
the reactions just described. Thus, any particle
can be destroyed with rate $\varepsilon_h \, h$, with $0 < \varepsilon_h < 1$, irrespective of 
whether neighboring sites are occupied or not, and a reversing of the annihilation process
yields the creation of particles with rate $\varepsilon_\lambda \, \lambda$, where $0 < \varepsilon_\lambda < 1$.
It is important to note that even with these expanded reaction schemes our systems are still
characterized by non-equilibrium steady states, as long as $\varepsilon_h , \varepsilon_\lambda < 1$.
In the following we set $\varepsilon_h = \varepsilon_\lambda = \varepsilon$ and focus on the
case $\varepsilon \ll 1$.
For the convenience of the reader we summarize the different reaction schemes in Table \ref{table1}.

%%%%%%%%%%%%%%%%%%%%%%%%%%%%%%%%%%%%%%%%%%%%%%%%%%%%%%%%%%%%%
\begin{table}[thb]
\begin{tabular}{|c|c|}
\hline
$M_{2'}$ & $M_n$ \\
\hline
$A+A\overset{\lambda}{\underset{\varepsilon_\lambda\lambda}{\rightleftarrows}} 0+A$ &
$nA \overset{\lambda}{\underset{\varepsilon_\lambda\lambda}{\rightleftarrows}}  n 0 $ \\
$0 \overset{h}{\underset{\varepsilon_hh}{\rightleftarrows}} A $ &
$0 \overset{h}{\underset{\varepsilon_hh}{\rightleftarrows}} A$ \\
\hline
\end{tabular}
\caption{The different reaction schemes discussed in this work. 
The back-reactions are taking place with rates $\varepsilon_hh$ and $\varepsilon_\lambda\lambda$,
with $0 < \varepsilon_h < 1$ and $0 < \varepsilon_\lambda < 1$.} \label{table1}
\end{table}
%%%%%%%%%%%%%%%%%%%%%%%%%%%%%%%%%%%%%%%%%%%%%%%%%%%%%%%%%%%%%

As transport process we consider the Partially Asymmetric Simple Exclusion Process (PASEP) where 
every configuration is reversible. In the Total Asymmetric Simple Exclusion Process (TASEP) in an open
one-dimensional system, particles that are fed into the system at, say, the left end with rate $\alpha$ 
can leave the system at the right end with rate $\beta$. Inside the system particles can only jump to a right neighboring
site, provided that site is not occupied. Obviously, microscopic reversibility is always broken for this model.
In the PASEP, however,
particles can always jump in both directions, provided that the chosen site is empty,
with the probability for a jump to the right being $p$, whereas the probability for a jump to the
left is $1-p$.
In addition, particles can exit the system at the left with probability 
$\varepsilon_\alpha \alpha$ (with $0 < \varepsilon_\alpha< 1$) and
enter at the right with probability $\varepsilon_\beta \beta$ (with $0 < \varepsilon_\beta < 1$). 
It follows that microscopic
reversibility is always fulfilled. In order to keep the number of parameters as small as possible,
we set $\varepsilon_\alpha = \varepsilon_\beta = \varepsilon$, with $\varepsilon \ll 1$. 

\section{Methods}

Whereas our main focus in the following will be on the large deviation function of the steady-state entropy production 
in the long time limit, we will also briefly discuss the entropy production at short times.

Focusing on our lattice models, let us
consider a path in configuration space $C_0 \longrightarrow C_1 \longrightarrow \cdots \longrightarrow
C_{M-1} \longrightarrow C_M$ that starts at some configuration $C_0$ and ends at some configuration $C_M$
after $M$ diffusion or reaction steps. Every configuration $C_i$ is uniquely characterized by the occupation numbers
(being 0 or 1) of all lattice sites. We will denote by $P_S(C_i)$ the probability to find
the configuration $C_i$ in the steady state. With every step $i$ leading from configuration $C_{i-1}$ to
configuration $C_i$ we associate a time increment $\tau_i$ given by
\begin{equation} \label{eq:taui}
\tau_i = \frac{1}{\sum_j \omega(C_{i-1} \longrightarrow \widetilde{C}_j)}
\end{equation}
where $\omega(C_{i-1} \longrightarrow \widetilde{C}_j)$ is the rate with which we go from configuration $C_{i-1}$ to 
any other accessible configuration
$\widetilde{C}_j$. The sum in the denominator is thereby a sum over all configurations $\widetilde{C}_j$
(including the configuration $C_i$ in which the system will be at step $i$) that can be reached from the configuration
$C_{i-1}$ through diffusion or reaction. With that we can assign a total time
\begin{equation} \label{eq:tau}
\tau = \sum\limits_{i=1}^M \tau_i
\end{equation}
to our trajectory in configuration space.

Along the same trajectory the total entropy production is given by \cite{Sei08}
\begin{eqnarray} \label{eq:ent_prod}
s_{tot} & = & \ln \frac{P_S(C_0)}{P_S(C_M)} + \ln \prod\limits_i \frac{\omega(C_{i-1} \longrightarrow C_i)}{
\omega(C_i \longrightarrow C_{i-1})} \nonumber \\
& = &  \ln \frac{P_S(C_0)}{P_S(C_M)} + s_m
\end{eqnarray}
where $s_m$ is the entropy produced in the particle bath connected to our system. The boundary term,
$\ln \frac{P_S(C_0)}{P_S(C_M)}$, which can be neglected  in the long time limit, needs to be included when
investigating fluctuation relations of the entropy production at short times.

It is worth pointing out that in Eq. (\ref{eq:ent_prod}) the transition rates from configuration $C_{i-1}$
to $C_i$ and from configuration $C_i$ to $C_{i-1}$ are showing up together. This is the reason why we have added 
the back-reactions to our reaction schemes.

In Section IV.A we will discuss the mean entropy production in the short time regime as a function of time $\tau$.
For very small $\tau$, i.e. values of $\tau$ where only few configurations are visited, 
this can be done in an exact way, where, starting from all possible initial states, all possible
trajectories are determined and the corresponding entropy changes are recorded \cite{Dor09,Dor09b}. This allows
us to determine in a numerically exact way the probability distribution of the total entropy production,
$p(s_{tot},\tau)$, after time $\tau$. 
For larger (but still small)
values of $\tau$, we can compute the mean entropy production through standard Monte Carlo sampling.
All the data discussed in the following are numerically exact data, with the exception of the data
shown in Fig. \ref{fig2} which have been obtained by Monte Carlo simulations.

With the definition (\ref{eq:ent_prod}) of the total entropy production the probability distribution
$p(s_{tot},\tau)$ obeys a detailed fluctuation theorem for any time $\tau$:
\begin{equation} \label{dft}
p(s_{tot},\tau)/p(-s_{tot},\tau) = \exp \left( - s_{tot} \right)~.
\end{equation}
Initially \cite{Eva93,Gal95,Kur98,Leb99} this relation was shown to be valid in the long-time limit when
only $s_m$ is considered. If one adds the boundary term, as done in Eq. (\ref{eq:ent_prod}), then
the theorem (\ref{dft}) holds true for any finite time interval $\tau$ \cite{Sei05}.

In the following much emphasis will be put on the large deviation function of the entropy production
(also called the rate function)
\begin{equation} \label{large_dev}
\chi(\sigma) = \lim_{\tau \to \infty}  \left[ - \frac{1}{\tau} \ln p(s_m,\tau) \right]
\end{equation}
which describes the asymptotic large fluctuations of the entropy production (in that limit we can neglect the boundary
term in (\ref{eq:ent_prod})):
\begin{equation}
p(s_m,\tau) \sim \exp \left( - \chi(\sigma) \tau \right)
\end{equation}
with the normalized entropy production rate
\begin{equation}
\sigma = s_m /\left( \tau \left< \dot{s}_m \right> \right)~.
\end{equation}
Here $\left< \dot{s}_m \right> = \left< \frac{ds_m}{d\tau} \right>$ is the mean entropy production rate.
In order to compute the large deviation function, we proceed as in Ref. \cite{Leb99} and
consider the generating function 
\begin{equation}
\left< \exp \left[-\mu s_m(\tau) \right] \right>  = \int ds_m \exp\left[ -\mu s_m(\tau) \right] p(s_m,\tau)
\end{equation}
of $s_m(\tau)$. The time evolution of the generating function is given by the linear operator
\begin{eqnarray} \label{eq:Lmu}
L_\mu & = & - \left[ \sum\limits_j \omega(C_j \longrightarrow C_i) e^{-\mu \ln\frac{\omega(C_j\to C_i)}{\omega(C_i\to C_j)}} - r(C_i)
\right] \nonumber \\
& = & - \left[ \sum\limits_j \omega(C_j \longrightarrow C_i)^{1-\mu} \omega(C_i \longrightarrow C_j)^{\mu} 
- r(C_i) \right] \nonumber \\
\end{eqnarray}
where $r(C_i) = \sum\limits_j \omega(C_i \longrightarrow C_j)$. 
%Expressing the generating function in its decomposition of eigenfunctions $f_i(\mu)$ of $L_\mu$, the lowest eigenvalue will dominate the long time behavior. 
The spectrum of $L_\mu$ is strictly positive for $\mu\neq0$, whereas for $\mu=0$ the lowest eigenvalue is zero and characterizes
the steady state of the system. 
%For this reason we name the smallest eigenvalue $\nu(\mu)$ defined as 
%\begin{equation}
%L_\mu f_0(\mu) = \alpha(\mu) f_0 (\mu)~.
%\end{equation}
In the long time limit the lowest eigenvalue $\nu(\mu)$ can equivalently be determined as
\begin{equation}
\lim_{\tau \to \infty}  \left[ - \frac{1}{\tau} \ln \left< \exp \left[-\mu s_m(\tau) \right] \right> \right] = \nu(\mu)~.
\end{equation}
For our finite systems we determine the eigenvalue $\nu(\mu)$ through the numerically obtained spectrum of the operator 
$L_{\mu}$ which is expressed by a finite matrix. This is done using standard algorithms.

This lowest eigenvalue allows us to obtain expressions for various quantities of interest. Thus the
mean entropy production rate is given by
\begin{equation} \label{eq:mepr}
\left< \dot{s}_m \right> = \left. d\nu(\mu)/d\mu \right|_{\mu = 0}~,
\end{equation}
whereas the large deviation function is given by the Legendre transform
\begin{equation} \label{eq:ldf}
\chi(\sigma)=\max_{\mu}\{\nu(\mu)-\langle \dot{s}_m\rangle \sigma \mu \}.
\end{equation}
The fluctuation theorem \cite{Eva93,Gal95,Kur98,Leb99} is revealed through a
symmetry relation for the large deviation function,
\begin{equation} \label{eq:chi_FT}
\chi(-\sigma) = \chi(\sigma) + \langle \dot{s}_m\rangle \sigma~,
\end{equation}
or, on the level of the eigenvalue, through the equivalent symmetry
\begin{equation} \label{eq:nu_FT}
\nu(\mu) = \nu(1-\mu)~.
\end{equation}
%Standard algorithms are employed for this procedure. 
%Determining the lowest eigenvalue for a large range of linear operators for the same system with the same set of reaction rates allows us to discuss and and further analyze its functional dependence on $\mu$.

\section{Entropy production}

\subsection{Entropy production at short times}
Before discussing the long-time behavior of the entropy production, which is the main focus of
the paper, we briefly discuss some properties of the entropy production in the short time limit. As an example,
we discuss the reaction-diffusion models, but corresponding results are obtained for the transport processes.

\begin{figure}[h]
\centerline{\epsfxsize=4.50in\ \epsfbox{
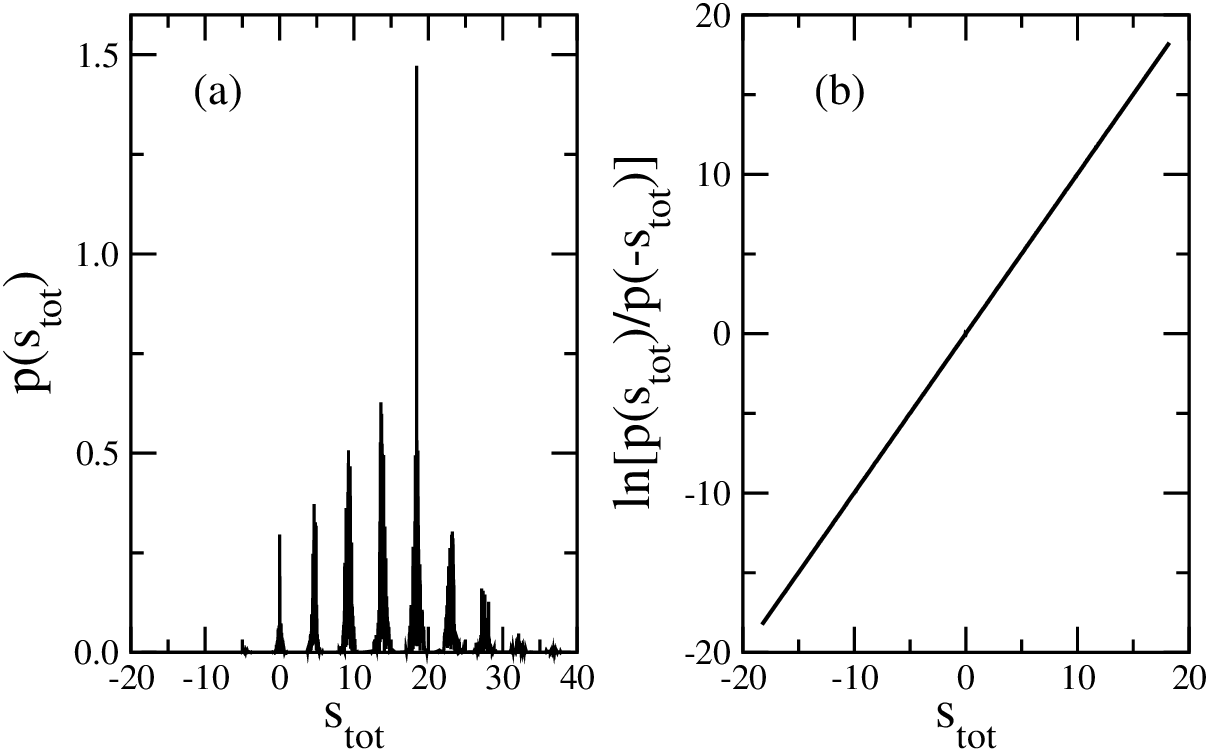}}
\caption{(a) Probability distribution for the total entropy production and (b) corresponding fluctuation theorem 
for model $M_3$ on a lattice with $N= 8$ sites. The values of the parameters are $M = 6$, $D = 1$, $\varepsilon
= 0.01$, $\lambda = 1$, and $h = 0.5$. 
}
\label{fig1}
\end{figure}

For all studied cases the probability distribution $p(s_{tot})$ for the total entropy production at short times 
is very irregular, as shown in Fig. \ref{fig1}a for the 
model $M_3$. Similar irregular shapes, dominated by prominent peaks, 
also show up when computing a related quantity, the so-called
driving entropy production, in systems that are driven out of stationarity through a change of system parameters
\cite{Dor09,Dor09b,Esp07}. 
%As discussed in \cite{Dor09b} the peaks are mainly due to trajectories in configuration
%space that involve a large number of diffusion steps and only few reactions. 
The details of the
probability distributions depend on the microscopic details of the studied systems, especially on the values
of the diffusion and reaction rates. Still, even though the probability distributions are very irregular
and specific for the studied systems, we recover in all cases the celebrated fluctuation theorem (\ref{dft}), see
Fig. \ref{fig1}b, provided that the boundary terms are included in the definition of $s_{tot}$.

%%%%%%%%%%%%%%%%%%%%%%%%%%%%%%%%%%%%%%%%%%%FIG 1.%%%%%%%%%%%%%%%%%%%%%%%%%%%%%%%%%%%%%%%%%%%%%%%%%%%%%%
%%%%%%%%%%%%%%%%%%%%%%%%%%%%%%%%%%%%%%%%%%%FIG 1.%%%%%%%%%%%%%%%%%%%%%%%%%%%%%%%%%%%%%%%%%%%%%%%%%%%%%%

\begin{figure}[t]
\centerline{\epsfxsize=4.50in\ \epsfbox{
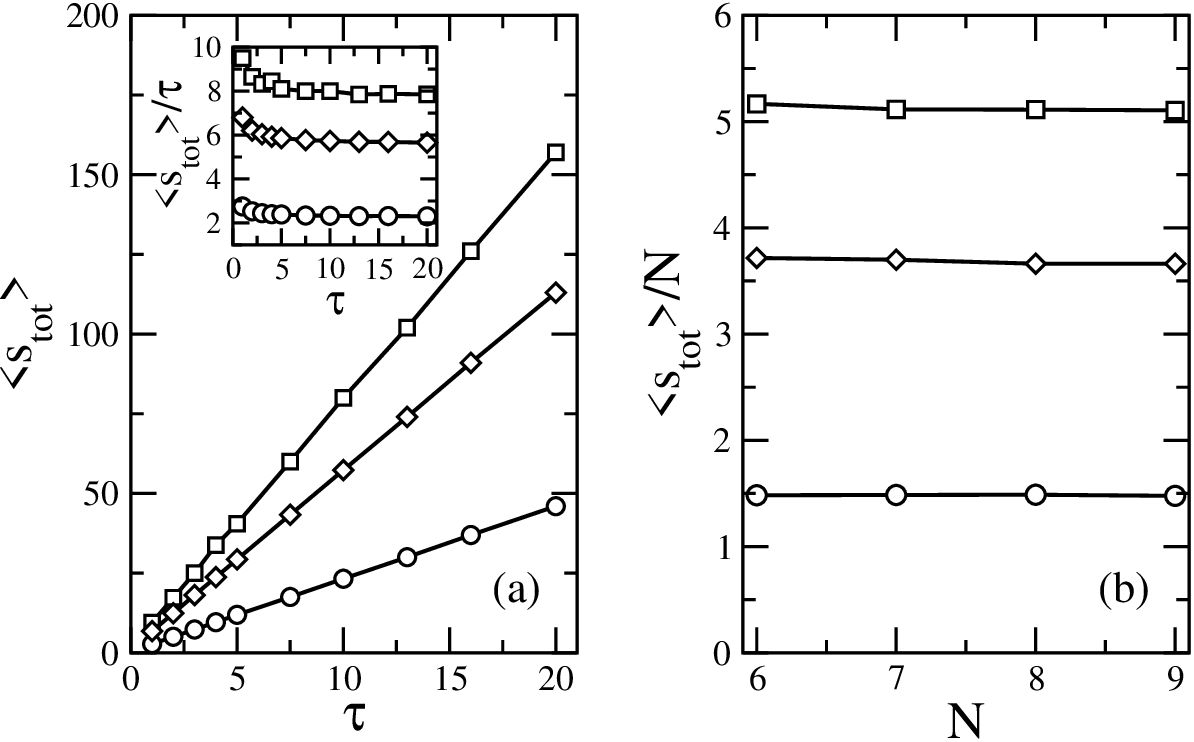}}
\caption{
Mean entropy production in the steady state as a function of the total measurement time $\tau$ (left) and system size $N$ (right)
for models $M_2'$ (circles), $M_2$ (squares) and $M_3$ (diamonds). The parameters are $D=1$, $h=0.5$, $\lambda=1$ and 
$\varepsilon=0.01$, with (a) $N=8$ and (b) $\tau=5$ for the right panel. Both the average entropy production per unit
time, see inset in (a), as well as the average entropy production per site reach rapidly a constant value.
}
\label{fig2}
\end{figure}

With the probability distribution of the total entropy production at hand, we can straightforwardly 
compute the mean value of entropy change $\left< s_{tot} \right>$ during some time interval $\tau$.
As shown for some cases in Fig.\ \ref{fig2}a, the total entropy production
rapidly yields a linear increase with time.
%This behavior is of course expected in the long time limit, but it is still remarkable that a linear
%increase prevails already for very short times. 
We also note, see Fig.\ \ref{fig2}b, that the entropy
change per site already becomes independent of the system size for very small systems. As a consequence of
these two observations we can restrict ourselves to small systems when discussing the long-time
properties in the following.

%%%%%%%%%%%%%%%%%%%%%%%%%%%%%%%%%%%%%%%%%%%FIG 2.%%%%%%%%%%%%%%%%%%%%%%%%%%%%%%%%%%%%%%%%%%%%%%%%%%%%%%
%%%%%%%%%%%%%%%%%%%%%%%%%%%%%%%%%%%%%%%%%%%FIG 2.%%%%%%%%%%%%%%%%%%%%%%%%%%%%%%%%%%%%%%%%%%%%%%%%%%%%%%

\subsection{The long time limit}
As discussed in Section III the long-time properties of the entropy production are encoded in the 
eigenvalue $\nu(\mu)$ of the linear operator $L_\mu$. After having determined $\nu(\mu)$ we
can use that quantity in order to discuss the mean entropy production rate, see Eq. (\ref{eq:mepr}), as well
as the large deviation function for the entropy production, see Eq. (\ref{eq:ldf}).

\subsubsection{The mean entropy production rate}
\begin{figure}[h]
\centerline{\epsfxsize=4.50in\ \epsfbox{
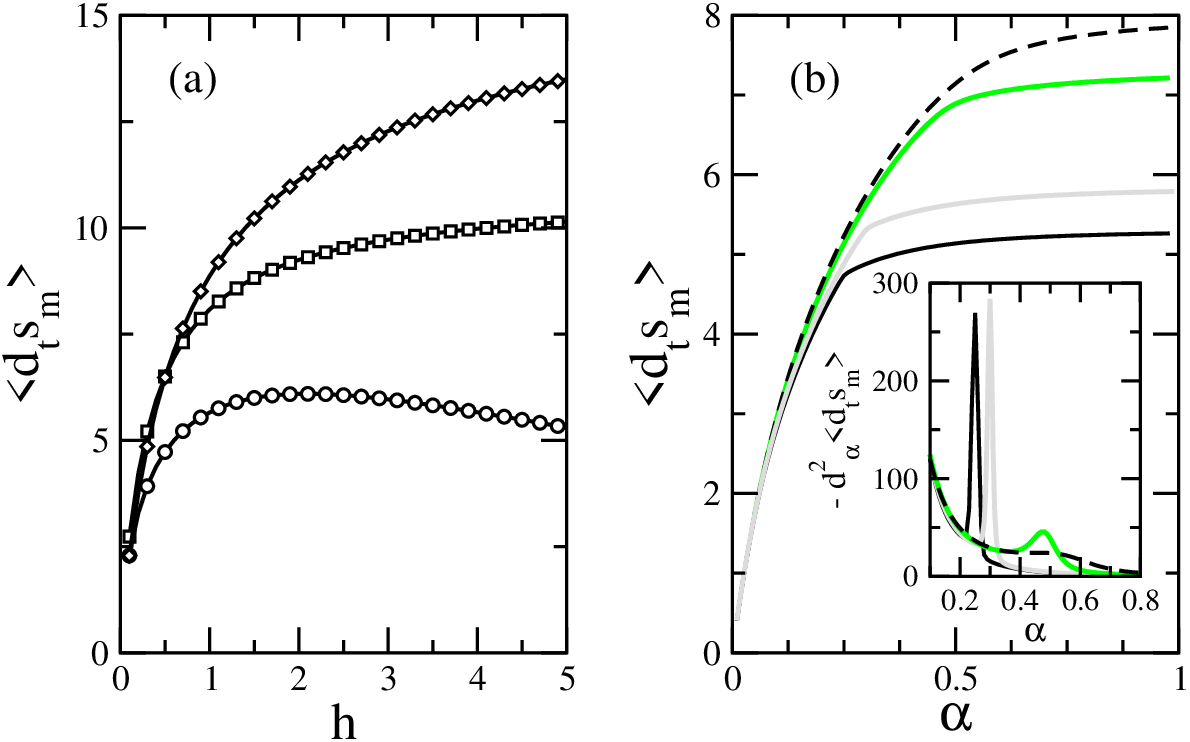}}
\caption{(Color online) Mean entropy production rate $\left< \dot{s}_m \right>$ for 
(a) reaction-diffusion systems $M_2'$ (circles), $M_2$ (squares) and $M_3$ (diamonds) as 
a function of the creation rate $h$ (values of other parameters: $N = 8$, $D = 0.5$, $\varepsilon = 0.01$,
with $\lambda = 5$ for model $M_2'$ and $\lambda = 0.1$ for models $M_2$ and $M_3$), and (b) 
the PASEP as a function of the entrance rate $\alpha$ (full black line: exit rate
$\beta = 0.25$; gray line: $\beta = 0.30$; dashed black line: $\beta = 0.5$; 
green (dark gray) line: $\beta = 0.75$) with $N=8$, $p= 0.95$
and $\varepsilon = 0.01$. Inset in (b): the negative of the second derivative of $\left< \dot{s}_m \right>$ with respect 
to $\alpha$ reveals the existence of a kink in the mean entropy production rate for $\alpha = \beta \lessapprox 0.46$.
For larger values of $\beta$ only a small peak remains, the position of which is $\alpha
\approx 0.47$, independent of the value of $\beta$.
}
\label{fig3}
\end{figure}

In Fig. \ref{fig3} we show examples of changes of the mean entropy production rate $\left< \dot{s}_m \right>$
for various systems when varying one of the system parameters. Fig. \ref{fig3}a illustrates
some of our results for the reaction-diffusion
systems where the creation rate $h$ is changed and all the other system
parameters are kept constant. Fig. \ref{fig3}b is devoted to the transport process with a varying
entrance rate $\alpha$.

Interestingly, $\left< \dot{s}_m \right>$ encodes important information on the physical properties
of the different steady states. Focusing first on the reaction-diffusion systems, we see that for 
model $M_2'$ the mean entropy production rate has a maximum around $h =1$ and decreases towards
zero for larger values of the creation rate. In stark contrast to this $\left< \dot{s}_m \right>$ increases
monotonously for the other two models $M_2$ and $M_3$. In order to understand this difference in behavior,
we refer the reader to the reaction schemes listed in Table \ref{table1}.
Noting that for
very large values of $h$ there is an increased probability 
for a free site to be surrounded by occupied neighboring sites, it follows that
the creation process $0\to A$ is then effectively identical to the process $A\to 2A$. This is, however, exactly the reversed
reaction to the annihilation process of model $M_2'$, such that this system approaches an equilibrium
system for very large values of $h$. Consequently, the mean entropy production rate vanishes in that
limit as no entropy production takes place in equilibrium. This is different for models
$M_2$ and $M_3$, as here the reaction $A\to 2A$ is not the direct back-reaction of the creation process, and
the systems remain out of equilibrium for any value of $h$.

Similar conclusions are also drawn when varying other system parameters. For example, when $\varepsilon \longrightarrow 1$,
all reactions become symmetric, and the system approaches equilibrium. Concomitantly, the mean entropy production
rate approaches zero in that limit.

It is worth noting that the behavior of the  mean entropy production rate is similar to that of a quantity 
derived from the stationary probability currents \cite{Zia06,Zia07,Pla10} that has been proposed as a possible measure for the 
non-equilibrium character of a system.
For example, both quantities show a non-monotonous change for model $M_2'$ when increasing $h$,
whereas their increase is monotonous for models $M_2$ and $M_3$ (see Fig.~2 in \cite{Dor09}).

%%%%%%%%%%%%%%%%%%%%%%%%%%%%%%%%%%%%%%%%%%%FIG 3.%%%%%%%%%%%%%%%%%%%%%%%%%%%%%%%%%%%%%%%%%%%%%%%%%%%%%%
%%%%%%%%%%%%%%%%%%%%%%%%%%%%%%%%%%%%%%%%%%%FIG 3.%%%%%%%%%%%%%%%%%%%%%%%%%%%%%%%%%%%%%%%%%%%%%%%%%%%%%%

Non-trivial information is also contained in the mean entropy production rate of the transport process.
As an example we show in Fig. \ref{fig3}b $\left< \dot{s}_m \right>$ as a function of the entrance rate $\alpha$
for a strongly asymmetric case with $p =0.95$. The striking feature in this figure is the appearance of 
kinks in the curves obtained for smaller exit rates, as for example  $\beta = 0.25$
or $\beta = 0.30$. These kinks, which are located at
$\alpha = \beta$, yield a sharp peak in the negative of the second derivative of $\left< \dot{s}_m \right>$ with
respect to $\alpha$, see inset of Fig. \ref{fig3}b. We checked that these kinks show up for all $\beta \lessapprox 0.46$ at the value 
$\alpha = \beta$. For larger values of $\beta$ the kink vanishes and only
a small peak remains in the second derivative (the peak still exists for $\beta = 0.75$ but is not visible 
on the scale of the inset in Fig. \ref{fig3}b). Interestingly, in this regime
the position of the peak is independent
of the value of $\beta$ and is given by $\alpha \approx 0.47$.

In order to understand this result we need to remind the reader of the phase diagram of the PASEP \cite{San94}.
For our choice of entrance and exit rates at the boundaries of the system 
%(entrance resp. exit rate at the
%left boundary: $\alpha$ resp. $\varepsilon \alpha$; entrance resp. exit rates at the right boundary:
%$\varepsilon \beta$ resp. $\beta$)
the phase diagram of the PASEP is very similar to that of the TASEP and contains a low density phase, a high density
phase and a maximum current phase. The phase boundary between the low and the high density phases, which is discontinuous 
and characterized by the coexistence of these two phases, is given by
$\alpha = \beta$, $\alpha < r_c$, where $r_c$ depends on the specific values of the different rates. In addition,
one has that the boundary between the low density (high density) phase and the maximum current phase is given by
$\alpha = r_c$ and $\beta > r_c$ ($\alpha > r_c$ and $\beta = r_c$). Insertion of the values of the rates used in Fig. \ref{fig3}b
into the exact expressions for the phase boundaries \cite{San94} yields the value $r_c = 0.45\overline{45}$. Comparing
this with the position of the kink in the mean entropy production rate for $\beta \lessapprox 0.46$,
we see that the kink coincides with the
discontinuous transition separating the low from the high density phase. In addition, for $\beta > r_c$ the extremum
of the second derivative of $\left< \dot{s}_m \right>$ close to $\alpha \approx r_c$ indicates the transition
separating the low density phase from the maximum current phase.

\subsubsection{Large deviation function for the entropy production}

In order to compute the large deviation function for the entropy production, we proceed as discussed in
Section III and determine the lowest eigenvalue $\nu(\mu)$ of the linear operator $L_\mu$, see Eq.
(\ref{eq:Lmu}). The large deviation function is then obtained through the Legendre transform (\ref{eq:ldf}).

\begin{figure}[h]
\centerline{\epsfxsize=4.50in\ \epsfbox{
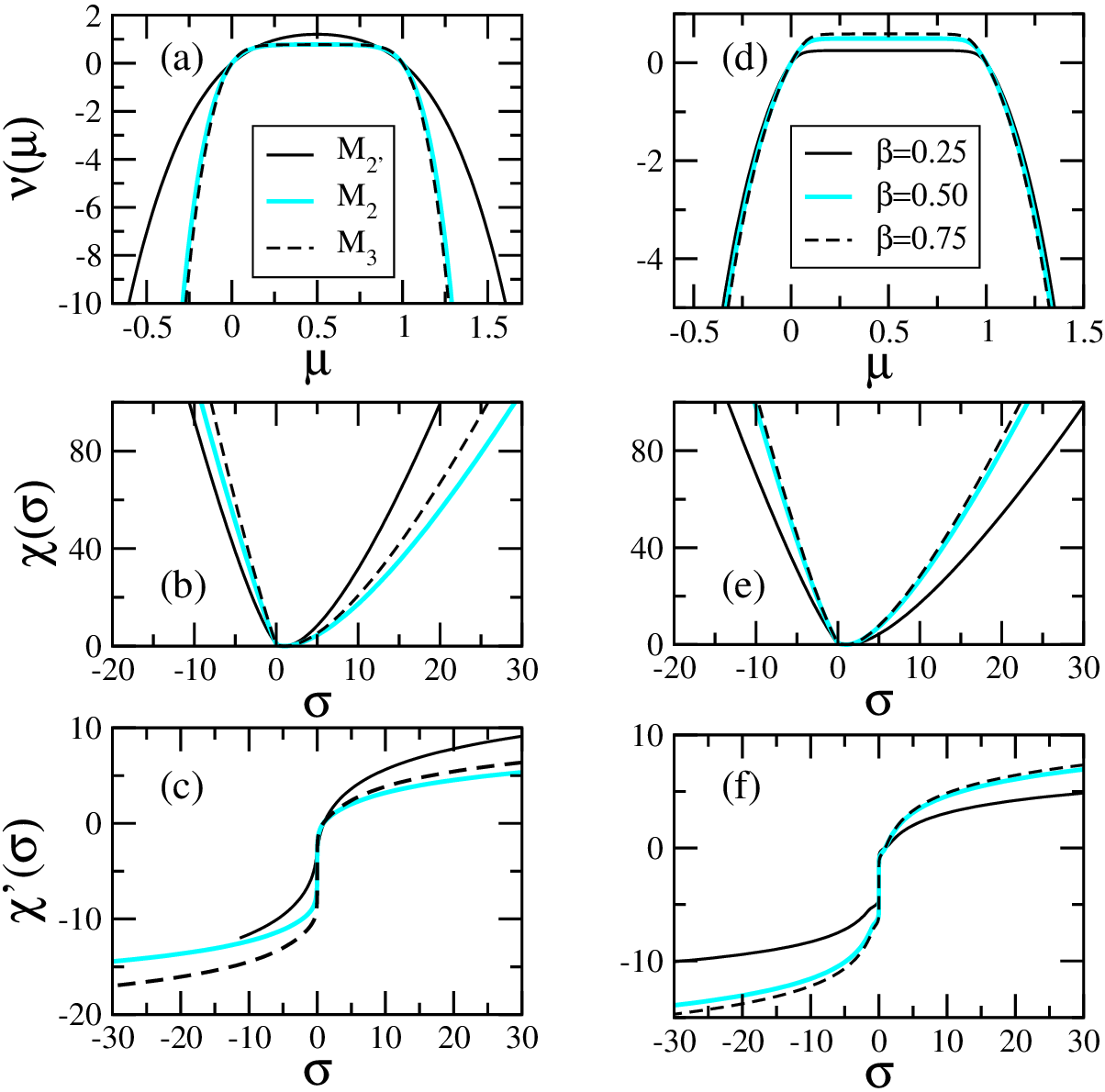}}
\caption{(Color online)
(a,d) The lowest eigenvalue $\nu(\mu)$ of the linear operator $L_\mu$, (b,e) the large deviation function for
the energy production $\chi(\sigma)$, and (c,f) the derivative of the large deviation function for (a,b,c)
the reaction-diffusion models shown in Fig. \ref{fig3}a, with $N = 8$, $h = 2$, $D = 0.5$, $\varepsilon = 0.01$,
$\lambda = 5$ for model $M_2'$ and $\lambda = 0.1$ for models $M_2$ and $M_3$, and (d,e,f) the transport processes discussed in
Fig. \ref{fig3}b, with $N=8$, $\alpha = 0.6$, $p= 0.95$
and $\varepsilon = 0.01$. In all studied cases, the large deviation function exhibits a kink at zero entropy production,
$\sigma = 0$
}
\label{fig4}
\end{figure}

Typical results obtained for our systems are shown in Fig. \ref{fig4} where the left column is devoted
to the reaction-diffusion systems and the right one to the PASEP. As expected, the eigenvalue $\nu(\mu)$
fulfills the relation (\ref{eq:nu_FT}) that follows directly from the fluctuation theorem, see Fig. \ref{fig4}a and
Fig. \ref{fig4}d. By construction, the large deviation function $\chi(\sigma)$ is zero at $\sigma = 1$. In addition,
plots of $\chi(\sigma)$, see Fig. \ref{fig4}b and Fig. \ref{fig4}e, reveal for all studied situations a kink
at zero entropy production $\sigma = 0$. This kink is best revealed when studying the derivatives of $\chi$ 
with respect to $\sigma$, as shown in Fig. \ref{fig4}c and Fig. \ref{fig4}e, yielding a step
at $\sigma = 0$ in the first derivative and a maximum at $\sigma = 0$ in the second derivative.

%%%%%%%%%%%%%%%%%%%%%%%%%%%%%%%%%%%%%%%%%%%FIG 4.%%%%%%%%%%%%%%%%%%%%%%%%%%%%%%%%%%%%%%%%%%%%%%%%%%%%%%
%%%%%%%%%%%%%%%%%%%%%%%%%%%%%%%%%%%%%%%%%%%FIG 4.%%%%%%%%%%%%%%%%%%%%%%%%%%%%%%%%%%%%%%%%%%%%%%%%%%%%%%

As stressed in the Introduction, a kink has also been observed recently in a study of the
steady-state entropy production for a single particle driven along a periodic potential \cite{Meh08}. 
The fact that a similar behavior is observed in our interacting many-body systems suggests that this
feature is a universal one. 

\begin{figure}[h]
\centerline{\epsfxsize=4.50in\ \epsfbox{
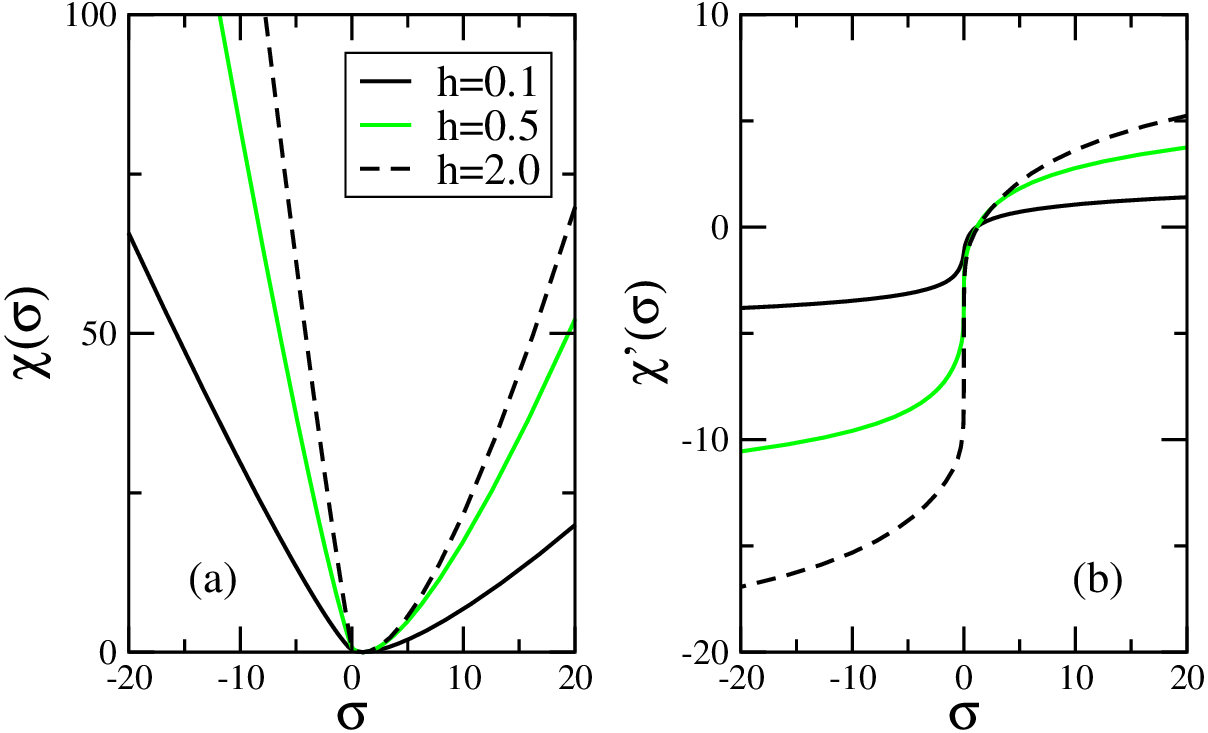}}
\caption{(Color online)
(a) The large deviation function of the entropy production and (b) its derivative for model $M_3$ and
different values of the creation rate $h$. The parameters used for this figure are $N = 8$, $D = 0.5$,
$\lambda = 0.1$, and $\varepsilon = 0.01$. For $\sigma = 0$ a kink shows up in the large deviation
function which is best illustrated by the rapid change in its derivative with respect to $\sigma$.
}
\label{fig5} 
\end{figure}

Before discussing the origin of the kink further, let us briefly pause and discuss how the large deviation
function and its derivative depend on the parameters of the system. As an example, we show 
in Fig. \ref{fig5} $\chi$ and $\chi'$ for the reaction-diffusion model $M_3$ for various values of the
creation field $h$. In our discussion of the mean entropy production rate we pointed out that for
this system $\left< \dot{s}_m \right>$ increases when $h$ increases, resulting in an enhanced non-equilibrium
character when taking $\left< \dot{s}_m \right>$ as a measure. Interestingly, the kink in the large deviation
function also gets more pronounced when $h$ increases, as witnessed by a larger amplitude of the step in the
derivative $\chi'$. This is the generally observed behavior for all our systems: when the system
parameters are varied in such a way that the mean entropy production rate increases (decreases), then the 
step height at $\sigma =0$ of the derivative of the large deviation function also increases (decreases).

%%%%%%%%%%%%%%%%%%%%%%%%%%%%%%%%%%%%%%%%%%%FIG 5.%%%%%%%%%%%%%%%%%%%%%%%%%%%%%%%%%%%%%%%%%%%%%%%%%%%%%%
%%%%%%%%%%%%%%%%%%%%%%%%%%%%%%%%%%%%%%%%%%%FIG 5.%%%%%%%%%%%%%%%%%%%%%%%%%%%%%%%%%%%%%%%%%%%%%%%%%%%%%%

%In order to make this qualitative statement more quantitative, we show in Fig.\ \ref{fig6} the step height,
%given bei the difference in slopes when approaching $\sigma =0$ from positive and negative values,
%$\chi'(0+) - \chi'(0-)$, for exactly the same cases as those used in Fig. \ref{fig3} to illustrate the
%behavior of the mean entropy production rate. Comparing both figures, an intimate
%relationship between $\chi'(0+) - \chi'(0-)$ and $\left< \dot{s}_m \right>$ is obvious: both quantities 
%have the same characteristics and both quantities reveal properties of the systems under study (behavior of
%the reaction-diffusion systems for large values of $h$, presence of a kink when crossing a phase boundary in
%the PASEP) in a similar way.
%
%%%%%%%%%%%%%%%%%%%%%%%%%%%%%%%%%%%%%%%%%%%%FIG 6.%%%%%%%%%%%%%%%%%%%%%%%%%%%%%%%%%%%%%%%%%%%%%%%%%%%%%%
%\begin{figure}[h]
%\centerline{\epsfxsize=3.00in\ \epsfbox{
%figure6.eps}}
%\caption{(Color online)
%Differences in the slopes of the large deviation function on both sides of the kink at $\sigma = 0$.
%The curves are obtained for the same cases as in Fig. \ref{fig3}.
%}
%\label{fig6}
%\end{figure}
%%%%%%%%%%%%%%%%%%%%%%%%%%%%%%%%%%%%%%%%%%%%FIG 6.%%%%%%%%%%%%%%%%%%%%%%%%%%%%%%%%%%%%%%%%%%%%%%%%%%%%%%

\begin{figure}[h]
\centerline{\epsfxsize=4.50in\ \epsfbox{
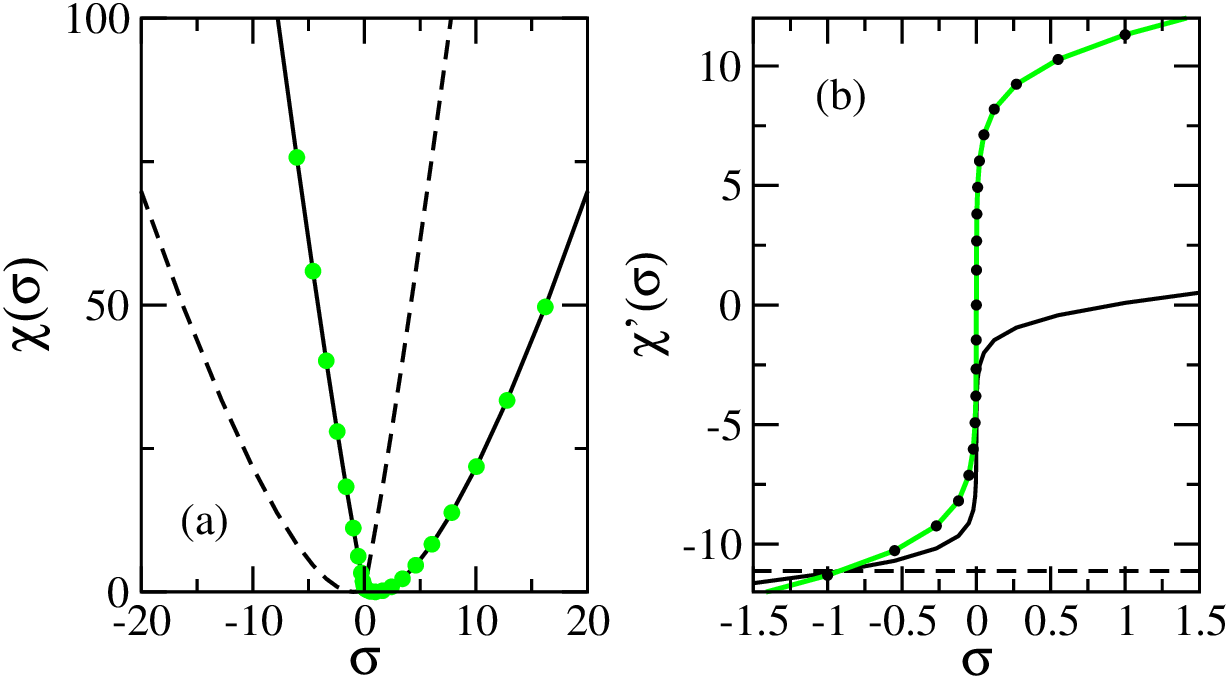}}
\caption{(Color online)
(a) Illustration of the symmetry relation (\ref{eq:chi_FT}) for model $M_3$ with $N= 8$, $D = 0.5$, $\varepsilon
= 0.01$, $h = 2$, and $\lambda = 0.1$. The full line is $\chi(\sigma)$, the dashed line is $\chi(- \sigma)$,
whereas the green (gray) circles follow from $\chi(-\sigma) - \sigma \left< \dot{s}_m \right>$. For the chosen
set of parameters we have that $\left< \dot{s}_m \right> = 11.1228$. (b)  Illustration of relations
(\ref{eq:chip_FT1})  and (\ref{eq:chip_FT2}) of the derivative $\chi'$ of the large deviation function.
Full black line: $\chi'(\sigma)$; dashed black line: $\chi'(\sigma) + \chi'(- \sigma)$;
green (gray) line: $\chi'(\sigma) - \chi'(- \sigma)$;
black circles: $2 \chi'(\sigma) + \left< \dot{s}_m \right>$. 
}
\label{fig6}
\end{figure}

In fact, the kink at $\sigma = 0$ in the large deviation function for the entropy production is a generic
feature and immediately follows from the fluctuation theorem  \cite{Eva93,Gal95,Kur98,Leb99}. To see this,
we remind the reader that the fluctuation theorem shows up on the level of the large deviation function
through the symmetry relation (\ref{eq:chi_FT}). 
Taking the derivative of that relation with respect to $\sigma$ yields the following relations between the
derivatives at both sides of $\sigma = 0$:
\begin{eqnarray}
\left. \frac{d \chi}{d \sigma} \right|_{\sigma = \sigma_0} + \left. \frac{d \chi}{d \sigma} \right|_{\sigma =
- \sigma_0} & = & - \left< \dot{s}_m \right> \label{eq:chip_FT1} \\
\left. \frac{d \chi}{d \sigma} \right|_{\sigma = \sigma_0} - \left. \frac{d \chi}{d \sigma} \right|_{\sigma =
- \sigma_0} & = & 2 \left. \frac{d \chi}{d \sigma} \right|_{\sigma = \sigma_0} + \left< \dot{s}_m \right>
\label{eq:chip_FT2}
\end{eqnarray}
valid for {\it any} value of the entropy production $\sigma_0$. In Fig. \ref{fig6} we illustrate for
model $M_3$ the relation (\ref{eq:chi_FT}) for the large deviation function as well as the relations 
(\ref{eq:chip_FT1})  and (\ref{eq:chip_FT2}) for its derivative.

%%%%%%%%%%%%%%%%%%%%%%%%%%%%%%%%%%%%%%%%%%%%FIG 6.%%%%%%%%%%%%%%%%%%%%%%%%%%%%%%%%%%%%%%%%%%%%%%%%%%%%%%
%%%%%%%%%%%%%%%%%%%%%%%%%%%%%%%%%%%%%%%%%%%%FIG 6.%%%%%%%%%%%%%%%%%%%%%%%%%%%%%%%%%%%%%%%%%%%%%%%%%%%%%%

The relations (\ref{eq:chip_FT1})  and (\ref{eq:chip_FT2}) completely characterize the kink. In fact, one
really should describe the observed feature as kink-like, as the derivative of $\chi$ is still continuous at $\sigma = 0$ 
even though its value changes rather dramatically when the entropy production approaches zero.
Eqs. (\ref{eq:chip_FT1})  and (\ref{eq:chip_FT2}) make obvious the connection between the kink and
the mean entropy production rate. It is the value of $\left< \dot{s}_m \right>$ that characterizes 
this intriguing behavior of the large deviation function
of the entropy production. A larger value of $\left< \dot{s}_m \right>$ yields a more pronounced feature and a
sharper and higher peak in the
second derivative of the large deviation function.

%Taking the limit $\sigma \longrightarrow 0+$, the symmetry
%relation immediately yields
%\begin{equation}
%\chi'(0-) = - \chi'(0+) - \left< \dot{s}_m \right>~.
%\end{equation}
%Consequently, a kink in $\chi$ appears at zero entropy production, characterized by the difference of
%slopes
%\begin{equation}
%\chi'(0+) - \chi'(0-) = 2 \chi'(0+) + \left< \dot{s}_m \right> = - 2 \chi'(0-) - \left< \dot{s}_m \right>~.
%\end{equation}

One might object that in some known cases the distribution of the entropy production 
has a nearly Gaussian form and that for these cases a kink-like behavior is not observed. 
An example can be found in Ref. \cite{Meh08} (see their Figures 1 and 3), where the large deviation function
for a particle driven along a periodic potential is nicely fitted to a parabola
\begin{equation} \label{eq:fit1}
\chi(\sigma) = \left( \langle \dot{s}_m \rangle /4 \right) \left( \sigma -1 \right)^2
\end{equation}
for small forces.
In a similar way, the largest eigenvalue seems to be well described by
\begin{equation} \label{eq:fit2}
\nu(\mu) = \langle \dot{s}_m \rangle \mu ( 1 - \mu )~.
\end{equation}
These results seem to indicate that close to equilibrium systems with a nearly Gaussian entropy production 
can fulfill the fluctuation theorem without exhibiting a kink-like feature in the large deviation function.

%%%%%%%%%%%%%%%%%%%%%%%%%%%%%%%%%%%%%%%%%%%FIG 7.%%%%%%%%%%%%%%%%%%%%%%%%%%%%%%%%%%%%%%%%%%%%%%%%%%%%%%
\begin{figure}[h]
\centerline{\epsfxsize=4.50in\ \epsfbox{
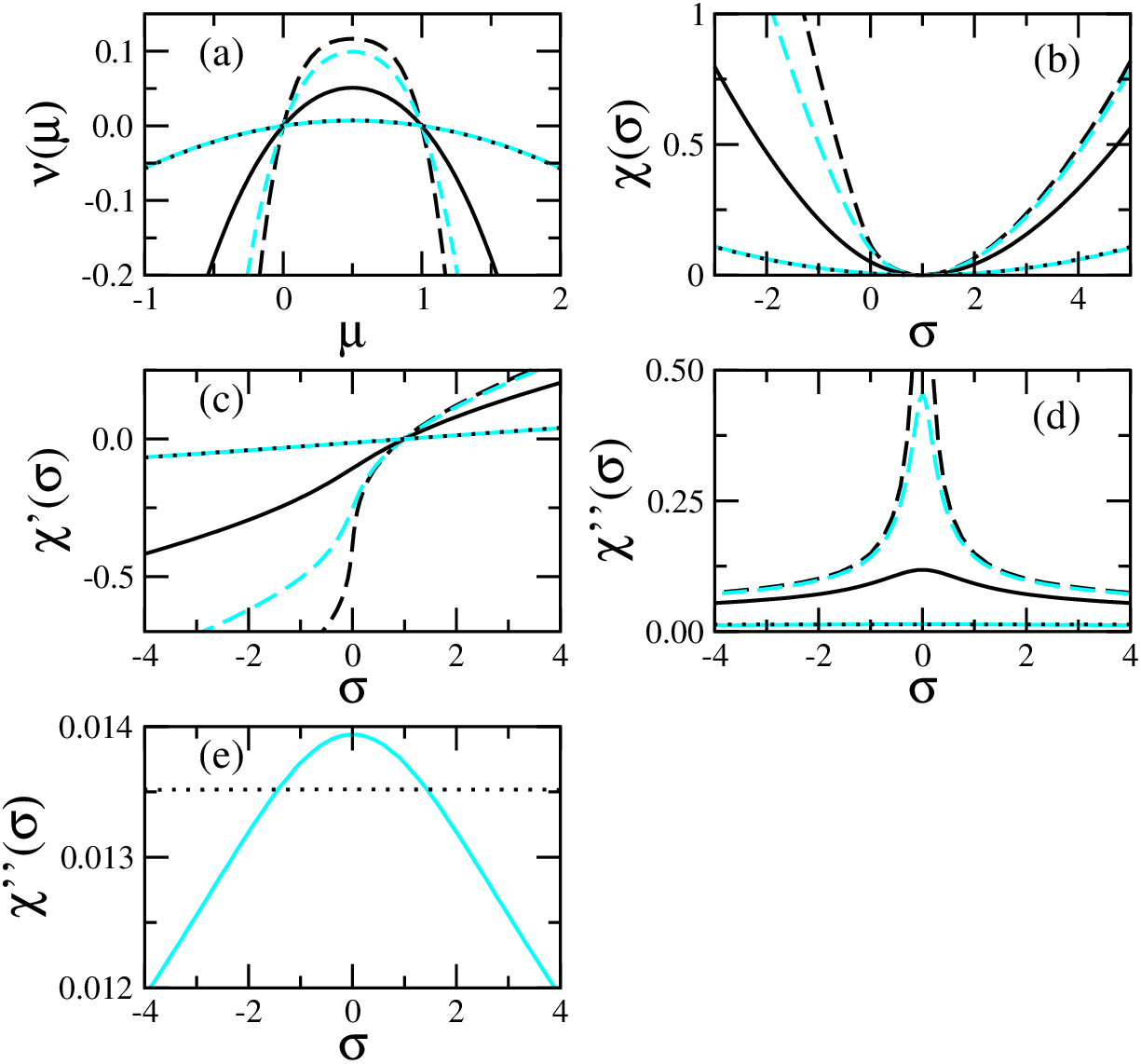}}
\caption{(Color online)
Discussion of the transport process with $\alpha = \beta = 0.5$ and $p = 0.5$. The different curves
correspond to various values of $\varepsilon$: 0.001 (dashed black line), 0.01 (dashed gray [magenta] line),
0.1 (full black line), and 0.5 (full gray [magenta] line). 
(a) Lowest eigenvalue $\nu(\mu)$ of the linear operator $L_\mu$, (b) large deviation function for the entropy 
function $\chi(\sigma)$, (c) first derivative of the large deviation function, (d) second derivative of the
large deviation function, and (e) second derivative of the large deviation function for $\varepsilon = 0.5$.
The dotted lines result from fitting the data for
$\varepsilon = 0.5$ to the quadratic forms (\ref{eq:fit1}) and (\ref{eq:fit2}). Systematic deviations between the data and the fit are
readily observed in $\chi''(\sigma)$.
}
\label{fig7}
\end{figure}
%%%%%%%%%%%%%%%%%%%%%%%%%%%%%%%%%%%%%%%%%%%FIG 7.%%%%%%%%%%%%%%%%%%%%%%%%%%%%%%%%%%%%%%%%%%%%%%%%%%%%%%

Our systems allow us to address this point in a systematic way. In fact, we always find for
systems that are close to equilibrium (as characterized by a small value of the mean entropy production
rate) that the entropy production approaches a nearly Gaussian form and that the kink seems to vanish.
This is illustrated in Fig. \ref{fig7} for the transport process with $\alpha = \beta = p = 0.5$ and various
values of $\varepsilon$. In this case the system approaches equilibrium when $\varepsilon$ increases, yielding
an equilibrium system for $\varepsilon = 1$. As we see from Fig. \ref{fig7}a and \ref{fig7}b, the kink-like feature
in the large deviation function
gets less and less pronounced when increasing $\varepsilon$ and seems to have vanished completely for
$\varepsilon = 0.5$. Indeed, in that case one can fit the largest eigenvalue and the large deviation function
to the quadratic forms (\ref{eq:fit2}) and (\ref{eq:fit1}), thereby obtaining an excellent agreement. However, a closer
look at the second derivative, see Fig. \ref{fig7}d and \ref{fig7}e, 
reveals that qualitatively the case $\varepsilon = 0.5$ does {\it not} differ from
the other cases: the second derivative exhibits a maximum at $\sigma = 0$ which is the signature of the kink-like
feature. Admittedly, this maximum is rather shallow for a system close to equilibrium, but it persists as long
as the mean entropy production rate does not vanish. Obviously, it is not good enough to simply fit the data to
the quadratic forms (\ref{eq:fit1}) and (\ref{eq:fit2}) and to conclude from an optically good fit that the
kink-like behavior is absent. The kink-like feature might no longer be easily detectable by simple inspection of $\chi$
in systems where the entropy production approaches a nearly Gaussian form,
but it still reveals itself in a maximum in $\chi''(\sigma)$. 

From our study, as well as from the observation of kink-like features in other systems fulfilling the fluctuation
theorem, we can therefore conclude that this behavior is generic for non-equilibrium systems.

\section{Conclusion}
Our study of interacting many-body systems has yielded two main results regarding the entropy production in the steady states of
non-equilibrium systems. 

On the one hand, we showed that the kink observed in the 
entropy production large deviation function at zero entropy production is generic and directly follows
from the fluctuation theorem. The well known relation for the large deviation function
(\ref{eq:chi_FT}) implies certain relations between the derivatives on both sides of $\sigma = 0$, see
Eq. (\ref{eq:chip_FT1}) and (\ref{eq:chip_FT2}). It follows that the kink is characterized
by the value of the mean entropy production rate in
the steady state.

On the other hand, our results indicate that the mean entropy production rate is a good measure for
quantifying the non-equilibrium character of a system (or, equivalently, the degree of violation of
detailed balance). Not only does it characterize the kink-like structure in the large deviation
function, it also directly reveals whether a system gets closer to be an equilibrium system when changing
some system parameters, as illustrated in Fig. \ref{fig3}a. In addition, the mean entropy production
rate also reveals the presence of dynamic phase transitions, as seen in our study of the PASEP, see
Fig.\ \ref{fig3}b.

Recently \cite{Zia06,Zia07,Pla10} it was proposed that the non-equilibrium character of a system
can be captured quantitatively by some global observables derived from the probability currents.
Probability currents and entropy production are of course intimately related. They are unique
to a non-equilibrium system and are absent
in equilibrium systems. They also show a remarkably similar dependence on system parameters.
As an example, the reader should compare our Fig.
\ref{fig3}a with Fig.~2 in \cite{Dor09} where these quantities are computed for
the one-dimensional reaction-diffusion systems discussed in this manuscript.
To our knowledge, however, it has not yet been studied
whether the observable derived from the stationary probability currents allows to 
probe the system for the presence of a phase transition. A more systematic comparison of the behavior
of that quantity with that of the mean entropy production rate seems to be needed in order
to further elucidate the commonalities and differences between these two quantities.

As a final remark we note that in the past quite some effort has been devoted to the computation and
understanding of the current large deviation function in driven systems. In our study we focused
on the large deviation function of the entropy production and characterized
its properties both for systems with (PASEP) and without (reaction-diffusion systems) particle
currents. The exact relation between these two large deviation functions in driven systems remains
to be better understood.\\

\begin{acknowledgments}
We thank Udo Seifert for suggesting this study and Uwe T\"{a}uber and Royce Zia for helpful discussions.
This work was supported by the US National
Science Foundation through DMR-0904999.
\end{acknowledgments}


\begin{thebibliography}{99}
\bibitem{footnote1} In most of these papers the authors did not discuss the kink explicitly.
However, in all cases the existence of the kink 
can either be derived from the exact expressions given for the large deviation functions or 
it can be inferred from the numerical data shown in various figures.
\end{thebibliography}
\end{document}